\pdfoutput=1
\documentclass[aps,prb,twocolumn,10pt]{revtex4-1}
\usepackage[utf8]{inputenc}
\usepackage{libertine}
\usepackage[libertine]{newtxmath}
\usepackage{mathtools,graphicx,microtype}
\usepackage[colorlinks=true,allcolors=blue]{hyperref}

\begin{document}

\title{Deformed Wigner crystal in a one-dimensional quantum dot}

\author{Yasha Gindikin and Vladimir A.\ Sablikov}

\affiliation{Kotel'nikov Institute of Radio Engineering and Electronics,
Russian Academy of Sciences, Fryazino, Moscow District, 141190,
Russia}

\begin{abstract}
The spatial Fourier spectrum of the electron density distribution in a finite 1D system and the distribution function of electrons over single-particle states are studied in detail to show that there are two universal features in their behavior, which characterize the electron ordering and the deformation of Wigner crystal by boundaries. The distribution function has a $\delta$-like singularity at the Fermi momentum $k_F$. The Fourier spectrum of the density has a step-like form at the wavevector $2k_F$, with the harmonics being absent or vanishing above this threshold. These features are found by calculations using exact diagonalization method. They are shown to be caused by Wigner ordering of electrons, affected by the boundaries. However the common Luttinger liquid model with open boundaries fails to capture these features, because it overestimates the deformation of the Wigner crystal. An improvement of the Luttinger liquid model is proposed which allows one to describe the above features correctly. It is based on the corrected form of the density operator conserving the particle number.
\end{abstract}

\maketitle

\section{Introduction}
One-dimensional (1D) quantum dots attract a great deal of attention as appealing model objects to study the effects of electron-electron (\textit{e-e}) interaction, which is principally important for 1D electron systems.~\cite{Kouwenhoven,Reimann} The interest to the many-electron state in bounded 1D systems is presently increased due to the recent progress in magnetotunnelling spectroscopy studies of such structures.~\cite{Fiete,Steinberg,Auslaender} As a consequence of \textit{e-e} interaction, electrons form the strongly correlated state, which is referred to as Luttinger liquid. Main distinctive features of a Luttinger liquid are the absence of fermionic quasiparticles, which manifests itself in the absence of the Fermi step in the momentum distribution function, the power-law behavior of spectral functions near the Fermi energy with interaction-dependent exponents,~\cite{Haldane,Voit} and Wigner-like correlations of electrons.~\cite{Schulz}

These properties were established for ideal, \textit{i.e.} boundless 1D systems. However, whether they hold for realistic mesoscopic systems of finite length is not obvious. Indeed, the presence of boundaries may strongly affect the electron state and excitations since 1D electron correlation functions are known to decay as a power of distance, hence there is no characteristic length. Many observations performed on mesoscopic structures containing a finite 1D system do not confirm the theoretical predictions made for ideal systems. Just recall the interaction-driven conductance renormalization~\cite{Kane} in an infinite 1D system that does not actually occur because of the contacts,~\cite{Tarucha,Maslov,Sablikov_contacts} the spin polarization~\cite{Berggren,Rokhinson} that should not exist according to the Mattis-Lieb theorem, the `0.7' anomaly,~\cite{0.7} many facts of the nonuniversality of conductance quantization,~\cite{Crook,Bird,Picciotto} which existing theories fail to explain, and so on.

It is commonly believed that electrons in a bounded 1D system form the Luttinger liquid. The boundary effect consists in the change of electron correlations near the ends which are described by additional boundary exponents. This conclusion was made in several theoretical works using the bosonization approach.~\cite{Fabrizio,Mattsson,Grioni} Unfortunately, this approach is based on a number of model assumptions, which are not well justified. These are (i) the extension of the linear electron spectrum to infinite negative energy, which results in a violation of the conservation laws,\cite{Sablikov,Starykh} (ii) the linearization of the electron spectrum, which can lead to a striking departure from the properties of a real system with quadratic dispersion relation,\cite{Khodas} and (iii) neglecting some parts of the $2k_F$ components of the electron densities in the \textit{e-e} interaction Hamiltonian.

The goal of our paper is to investigate the electron state in 1D quantum dots beyond model assumptions. For this purpose the exact diagonalization method is employed to calculate numerically the electron density distribution in a finite 1D wire and the distribution function of electrons over single-particle states. We have found that there are two unexpected features. First, the distribution function has a $\delta$-like singularity at the Fermi energy at the background of a smooth dependence on the energy. The second feature relates to the Fourier spectrum of the spatial distribution of the electron density. The Fourier spectrum has a step-like form at the wavevector $2k_F$. Above this threshold, the harmonics are absent or vanishing. These properties are universal in the sense that they do not depend on the \textit{e-e} interaction strength, interaction radius, wire length, mean electron density. We argue that these features are caused by the Wigner ordering of electrons. We compare the obtained results with the calculations carried out within the frame of the common Luttinger liquid approach to find that it fails to capture the above results. The Luttinger liquid theory also gives a singularity of the distribution function at the Fermi energy, but its form is incorrect. The Fourier spectrum of the electron density distribution turns out to have an incorrect singularity at $2k_F$ and not to vanish above this value. We have clarified that this discrepancy appears because the Luttinger liquid theory does not describe correctly the deformation of the Wigner crystal by the boundaries. The reason of this shortcoming lies in the violation of the particle number conservation within the bosonization approach used. We find a way to remedy the Luttinger liquid theory by introducing an improved expression for the electron density operator. The proposed approach allows one to describe correctly the above features of the distribution function and the Fourier harmonics of electron density.

The formulation of problem and exact diagonalization results are presented in section~\ref{eds}. The results of bosonization with zero boundary conditions are given in section~\ref{fg_results}. Section~\ref{interpret} contains a simple model of Wigner ordering, as well as the interpretation of the results.
Section~\ref{bpnc} includes the review of bosonization approach for the case of zero boundary conditions, the construction of the density operator and the calculation of the observables with the particle number conservation taken into account. Technical details pertaining to the exact diagonalization can be found in Appendix~\ref{meth}.

\section{Exact diagonalization study of 1D correlated state}
\label{eds}
\subsection{Statement of problem}
Consider $N$ spinless electrons in a 1D quantum box with zero boundary conditions for the many-electron wavefunction, 
\begin{equation}
\Psi |_{x=0}=\Psi |_{x=L}=0\;.
\end{equation}
The Hamiltonian is
\begin{equation}
H=-\frac{\hbar^2}{2m}\sum_i \frac{\partial^2}{\partial x_i^2}+\sum_{i>j}V(x_i-x_j)+\sum_i U(x_i)\;,
\label{Hamiltonian}
\end{equation}
where $V(x)$ is the \textit{e-e} interaction potential, and $U(x)$ is the potential of positively charged background, which is considered as jelly.

Exact diagonalization method~\cite{Hylleraas} reduces to finding the Hamiltonian matrix in the appropriate basis $\Phi_\mathbf{p}$ and solving the eigenvalue problem by the standard methods of computational linear algebra.~\cite{Golub} As a result we obtain the many-electron wave function $\Psi(x_1,..,x_N)$, expanded in this basis,
\begin{equation}
\Psi(x_1,..,x_N)=\sum_\mathbf{p} a_\mathbf{p}\Phi_\mathbf{p}(x_1,..,x_N)\;,
\label{expansion}
\end{equation}
and the spectrum. It is convenient to choose the basis functions in the form of the Slater determinants
\begin{equation}
\Phi_\mathbf{p}(x_1,..,x_N)=\dfrac{1}{\sqrt{N!}}
\begin{vmatrix}
\psi_{\alpha_1}(x_1)&\cdots&\psi_{\alpha_N}(x_1)\\
\vdots\\
\psi_{\alpha_1}(x_N)&\cdots&\psi_{\alpha_N}(x_N)
\end{vmatrix}\;,
\label{many_particle_wf}
\end{equation}
built from the eigenfunctions of the non-interacting single-particle Hamiltonian,
\begin{equation}
\psi_q(x)=\sqrt{\frac{2}{L}}\sin \frac{\pi qx}{L}\;.
\label{one_particle_wf}
\end{equation}
The quantum number $q \in \mathbb{N}$, labelling single-particle states, is analogous to momentum in translationally invariant systems, and will be called so for brevity. The many-particle state, labelled by the vector $\mathbf{p}=(\alpha_1,..,\alpha_N)$, is obtained by occupying the single-particle states $\psi_{\alpha_i}$, $i=1\dots N$. We adhere to the ordering convention that $\alpha_i<\alpha_k$ for $i<k$.

The Hamiltonian matrix elements are given in Appendix~\ref{meth}. The interaction potential is chosen in the form $V(x_1-x_2)=Ve^2/(2\epsilon d)\exp(-|x_1-x_2|/d)$, with $d$ being the interaction radius, $\epsilon$ the permittivity of the medium. This allows us to find the matrix elements analytically to cut the calculation time. Using such form of the potential is not essentially a limitation, because in our calculations we are able to vary the parameters $V$ and $d$ in a broad range to explore both the short-range and long-range (on the system length scale) interaction cases. 

Now let us express the observables we are going to find via the coefficients of expansion~(\ref{expansion}). The momentum distribution function is defined as
\begin{equation}
n(q)=\left\langle \Psi\right|c_q^+c_q\left|\Psi\right\rangle \;,
\end{equation}
where $c_q$ is electron destruction operator. Since $\left|\Phi_\mathbf{p}\right\rangle$ is the eigenvector of $c_q^+c_q$,
\begin{equation}
n(q)=\sum_\mathbf{p} |a_\mathbf{p}|^2\theta_\mathbf{p}^q\;,
\label{momentum}
\end{equation}
where the function $\theta_\mathbf{p}^q$ equals one if the many-particle state $\Phi_\mathbf{p}$ has the single-particle state $\psi_q$ occupied, and zero otherwise. 

The average value of the particle density operator
\begin{equation}
\rho=\sum_{i=1}^N\delta(x-x_i)
\label{pardenop}
\end{equation}
equals
\begin{equation}
\begin{split}
\rho(x)&=\left\langle \Psi\right|\rho\left|\Psi\right\rangle =\\
&N\int\Psi^*(x,x_2,..x_N)\Psi(x,x_2,..x_N)\,dx_2...dx_N\;.
\end{split}
\end{equation}
Using~(\ref{expansion}), one gets
\begin{equation}
\rho(x)=\sum_{\mathbf{p}_1,\mathbf{p}_2}a^*_{\mathbf{p}_1}a_{\mathbf{p}_2}\gamma_{\mathbf{p}_1\mathbf{p}_2}(x)\;,
\end{equation}
where for $\mathbf{p}_1=\mathbf{p}_2=(\alpha_1,..,\alpha_N)$
\begin{equation}
\gamma_{\mathbf{p}_1\mathbf{p}_1}(x)=\sum_{i=1}^N |\psi_{\alpha_i}(x)|^2\;,
\label{gamma1}
\end{equation}
and
\begin{equation}
\gamma_{\mathbf{p}_1\mathbf{p}_2}(x)=(-1)^{k_1+k_2}\psi^*_{\alpha_{k_1}}(x)\psi_{\alpha_{k_2}}(x)
\label{gamma2}
\end{equation}
for the case when $\mathbf{p}_1$ and $\mathbf{p}_2$ have only two different occupied states $\psi_{\alpha_i}$ in positions $k_1$ and $k_2$, respectively; $\gamma=0$ otherwise. The cosine Fourier-transform of the density is
\begin{equation}
\rho(q)=\int_0^L\rho(x)\cos\frac{\pi qx}{L}dx,\; q\in \mathbb{N}\;,
\label{cosine_Fourier}
\end{equation}
and sine Fourier-transform is zero, since according to Eqs.~(\ref{gamma1}),~(\ref{gamma2}), $\rho(x)$ contains only cosine harmonics.

\subsection{Results}
Below we present the results obtained for $N=12$ electrons. The system length is $L=333\,a_B$, the interaction radius $d=33\,a_B$, with $a_B$ being effective Bohr's radius, $V=3.6$. The corresponding value of the RPA parameter $r_s=(2na_B)^{-1}$ is 13.9, the estimate of the Luttinger liquid interaction parameter according to $g=(1+V_{q=0}/\pi \hbar v_F)^{-0.5}$ gives 0.3. 

\begin{figure}
\includegraphics[width=1.0\linewidth]{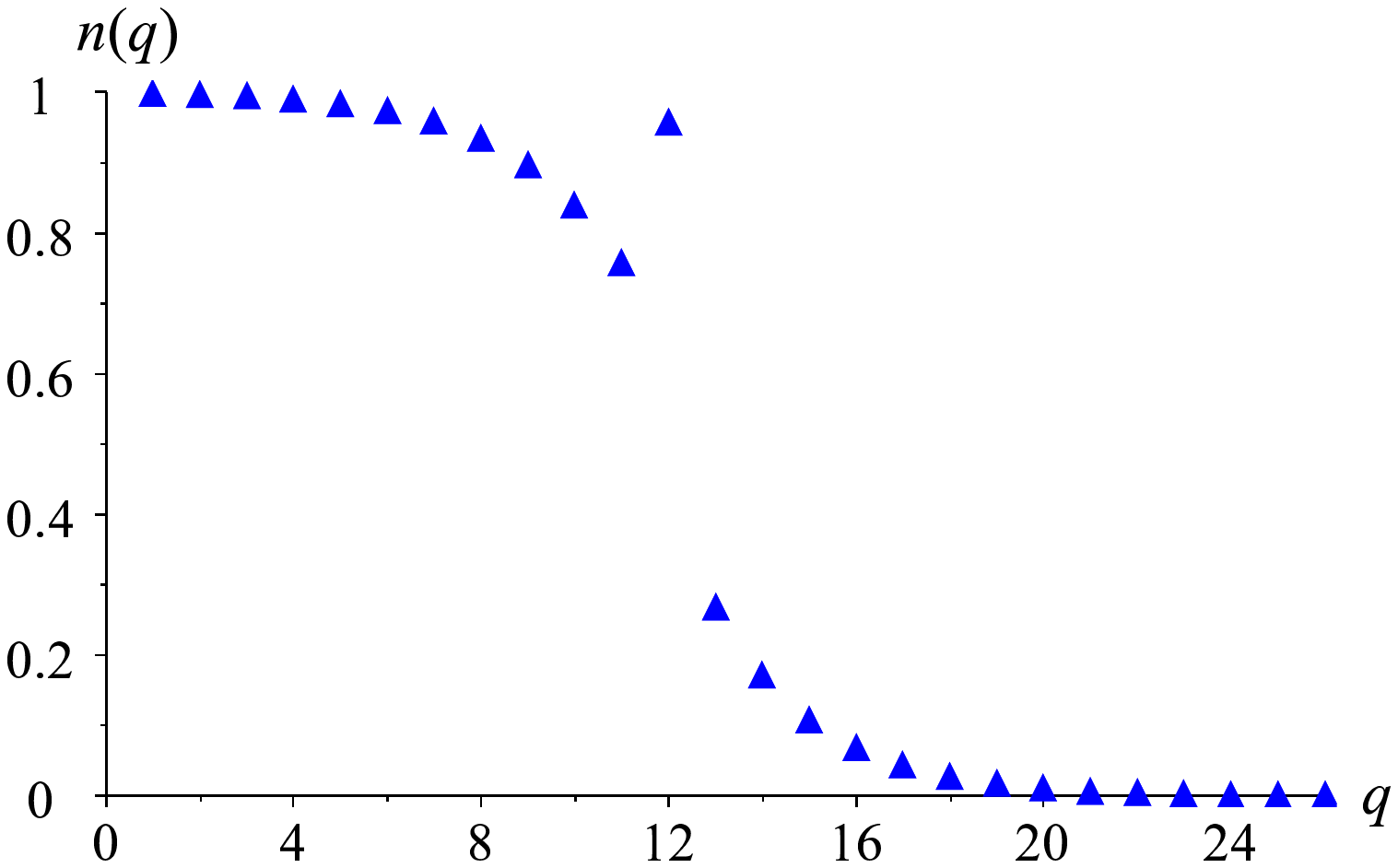}
\caption{Momentum distribution function, calculated for the system of $N=12$ electrons. The value of $q=12$ corresponds to $k_F$.}
\label{fig1}
\end{figure}

The distribution function $n(q)$ over the single-particle states $\psi_q$ is shown in Fig.~\ref{fig1}. Far from the Fermi surface, the form of the $n(q)$ curve is smoothed by the interaction, which is familiar from the Luttinger model. However, right at the Fermi surface there appears an unexpected $\delta$-type singularity, with the value of the distribution function being close to 1 at this point.

The presence of the singularity was checked by changing the number $N$ of electrons from 3 to 20, varying the length parameters and the interaction strength by the two orders of magnitude. The result proved to be perfectly stable against the change in the system parameters $(L,d,V,N)$. Hence, the $\delta$-singularity in the momentum distribution $n(q)$ at $q=k_F$ is a universal property of finite 1D systems. Its origin is explained in section~\ref{interpret}.

\begin{figure}
\includegraphics[width=1.0\linewidth]{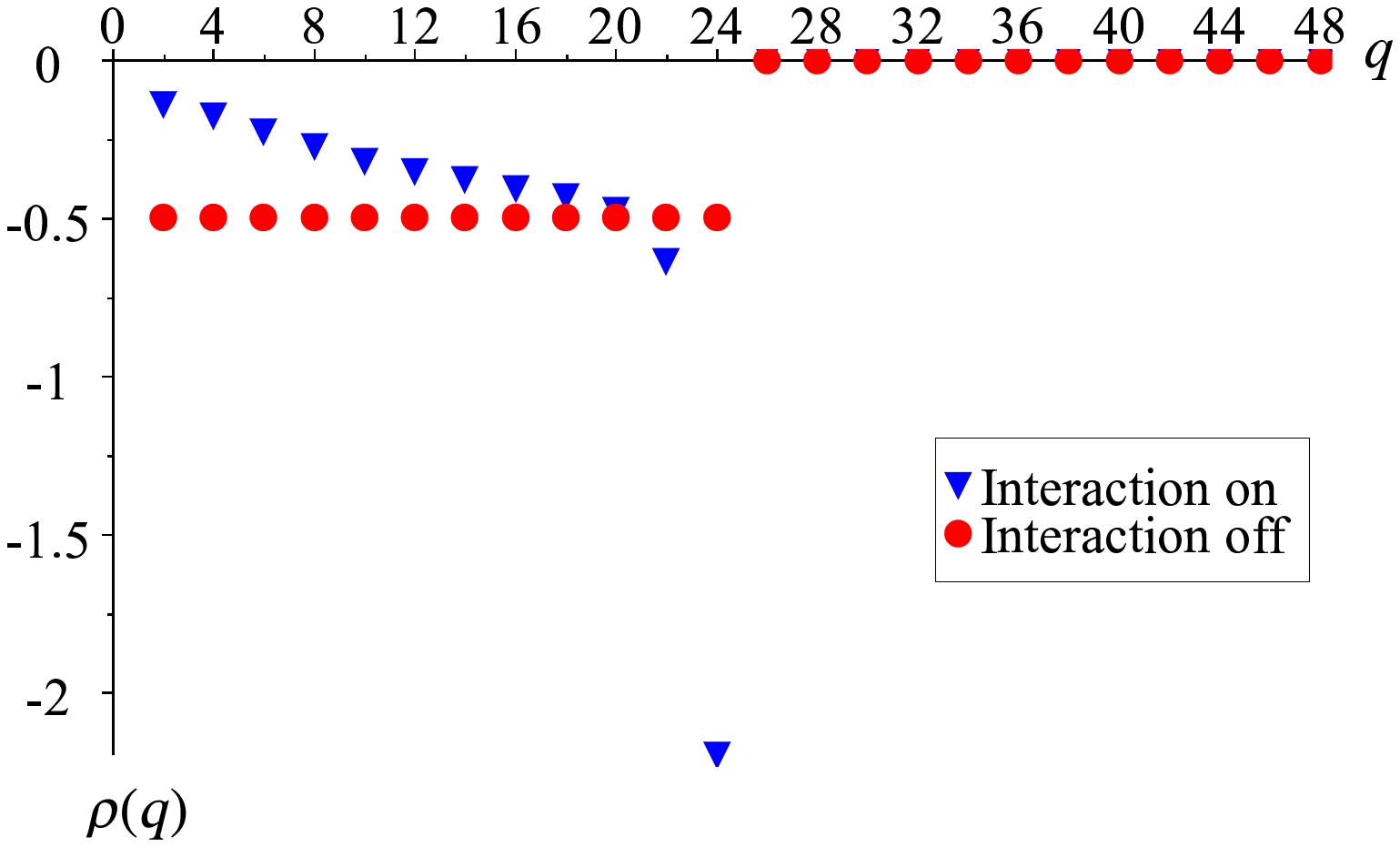}
\caption{The Fourier-transform of electron density. The value of $q=24$ corresponds to $2k_F$.}
\label{fig2}
\end{figure}

Electron density $\rho(x)$ in the ground state is an oscillatory function, the amplitude of which decays off the boundaries. To analyze the ordering in the electron state, consider the Fourier-transform of electron density $\rho(q)$, which is shown in  Fig.~\ref{fig2}. For comparison, the results for non-interacting electrons are also provided.

For free electrons, $\rho(q)$ has a step-like structure, with $\rho(q)=-0.5$ for $0<q\le 2k_F$, and $\rho(q)=0$ for $q>2k_F$. We emphasize that such threshold behavior holds even for a strongly interacting system, where $\rho(q)$ remains very close to zero for $q>2k_F$. Electron-electron interaction modifies the values of density harmonics only for $0<q\le 2k_F$.

The harmonics $\rho(q)$ for $0<q<2k_F$ are suppressed by the interaction. The $q=2k_F$ harmonic of the density is enhanced by interaction, reaching the values comparable to one half of the background density. This reflects the well-known fact that \textit{e-e} interaction leads to the strong electron correlations on the scale of average inter-particle distance, or in other words, to the Wigner-type ordering in the system.

The presented results are exact, since they are based on a precise many-particle wave function. In the next section we compare them to the results of the Luttinger liquid theory.

\section{Bounded Luttinger liquid theory}
\label{fg_results}
One of the most advanced analytical theories of a strongly correlated electron state in 1D quantum dots is a Luttinger liquid theory, based on the bosonization with zero (open) boundary conditions.~\cite{Fabrizio,Mattsson,Grioni} In this theory the spatial distribution of electron density and the distribution of electrons over single-particle states are expressed through the Green function of chiral fermions $G_+(x,y)=\left\langle \psi_+^+(y)\psi_+(x)\right\rangle$ via
\begin{equation}
n(q)=\frac{1}{2L}\int_{-L}^{L}dxdy\,G_+(x,y) e^{iq(y-x)}\;,
\label{mdf}
\end{equation}
and $\rho(x)= -e^{-2ik_Fx}G_+(-x,x)+ c.c.$ The Green function is
\begin{equation}
\begin{split}
&G_+(x,y)=\frac{1}{\beta L}\frac{(e^{\beta}-1)^{\frac{g^2+1}{2g}}}{2^{\frac{(g+1)^2}{4g}+1}}\left( e^{\beta}-e^{i\frac{\pi}{L}(x-y)}\right)\times\\
&\frac{\left[(\cosh \beta-\cos \frac{2\pi x}{L})(\cosh \beta-\cos \frac{2\pi y}{L})\right]^{\frac{g^{-1}-g}{8}} }{\left[\cosh \beta-\cos \frac{2\pi(x-y)}{L}\right]^{\frac{(g+1)^2}{4g}}\left[\cosh \beta-\cos\frac{2\pi(x+y)}{L}\right]^{\frac{1-g^2}{4g}}},
\label{green}
\end{split}
\end{equation}
with the dimensionless cutoff parameter $\beta\approx N^{-1}$. 

\begin{figure}
\includegraphics[width=0.9\linewidth]{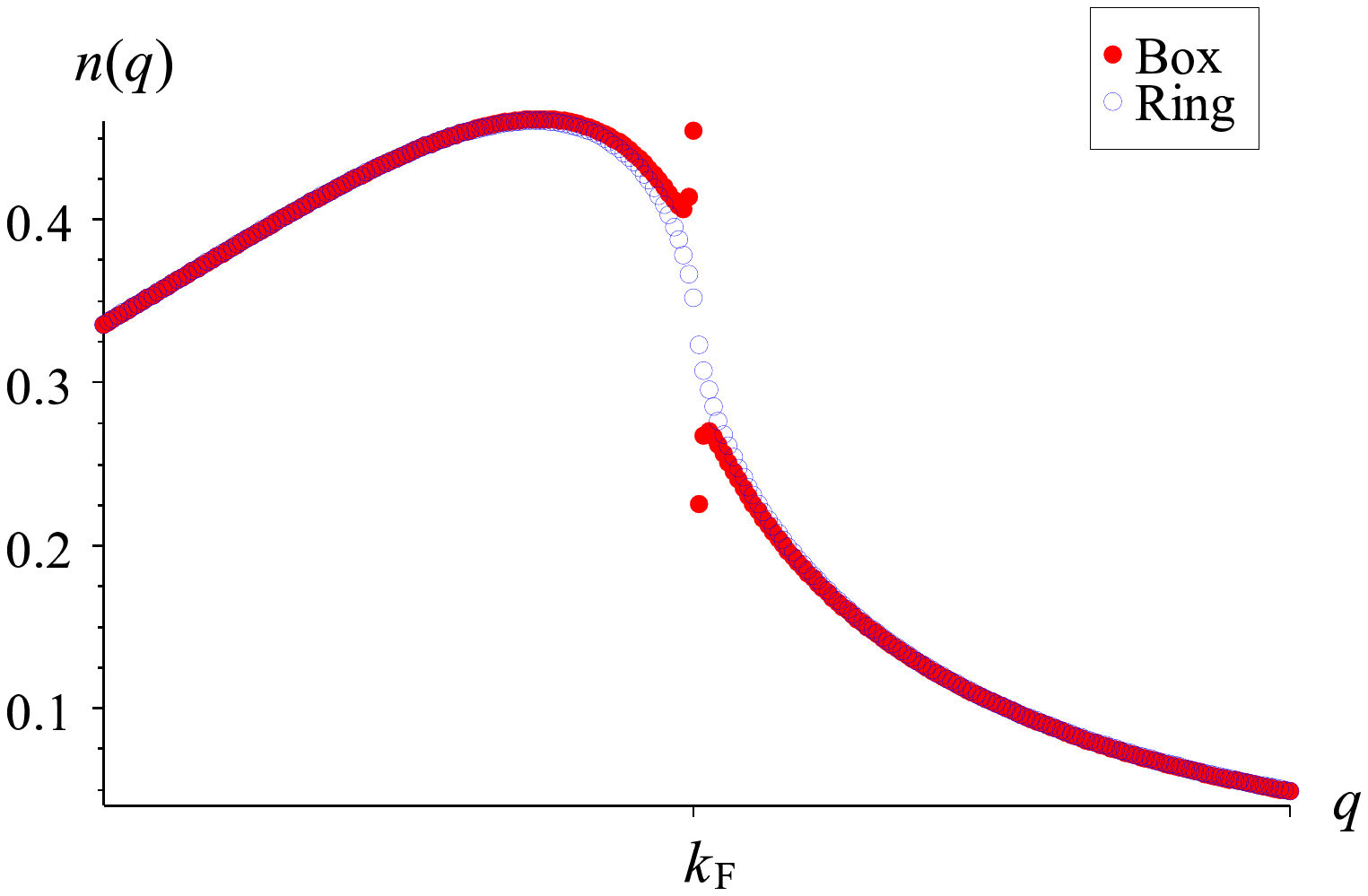}
\caption{Momentum distribution in the Luttinger model with zero (filled circles) and periodic (empty circles) boundary conditions, the interaction parameter $g=0.3$.}
\label{fig3}
\end{figure}
The momentum distribution function is presented in Fig.~\ref{fig3}. A comparison with exact diagonalization shows that this result is qualitatively incorrect. The momentum distribution function, calculated within bosonization, also has a singularity at $q=k_F$, but instead of a single $\delta$-peak at $q=k_F$, $n(q)$ deviates from a smoothed step in a finite band around $k_F$, where the form of the curve is close to the derivative of the $\delta$-peak.~\cite{footnote1}

\begin{figure}
\includegraphics[width=0.9\linewidth]{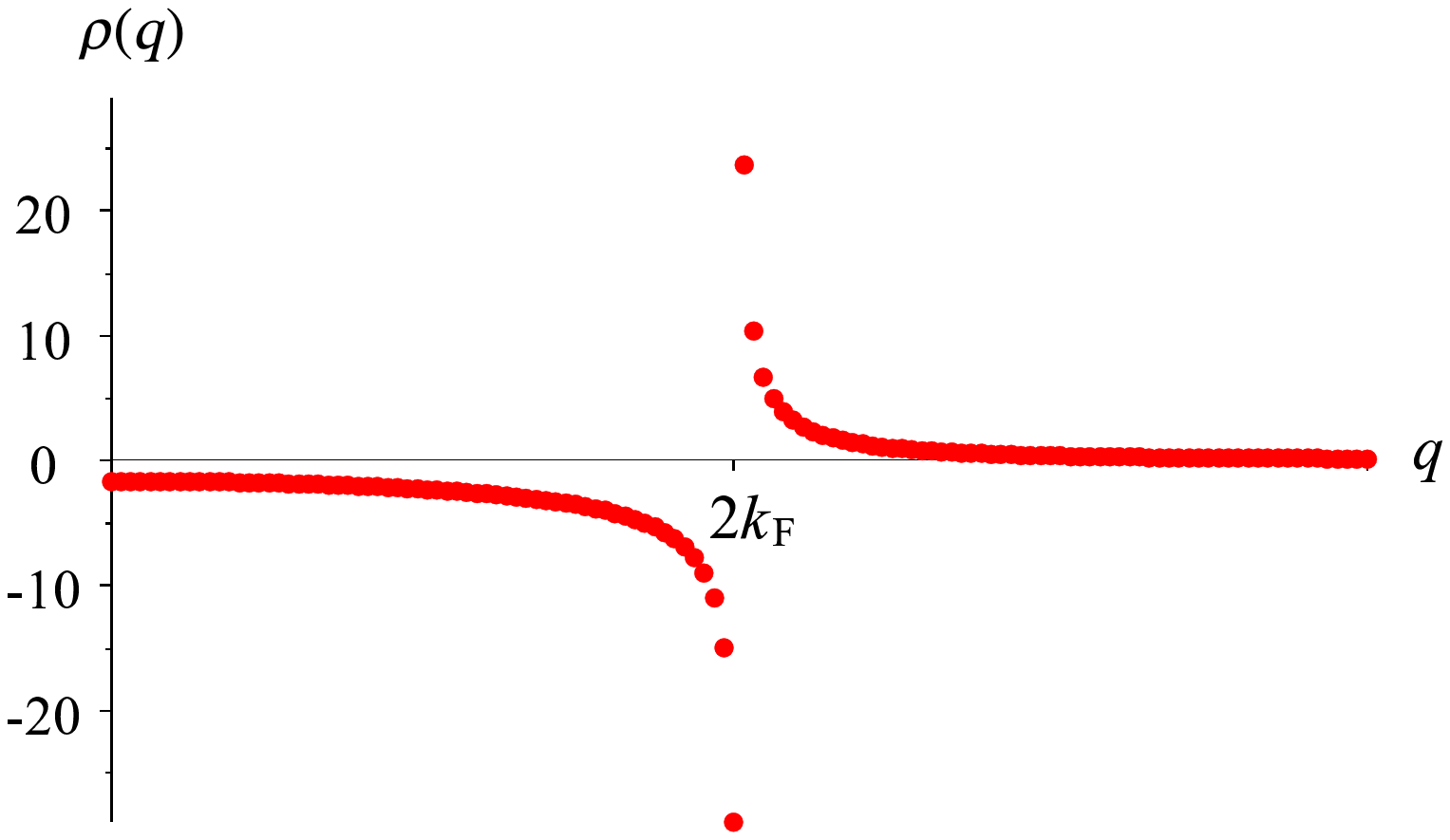}
\caption{The Fourier-transform of electron density in the Luttinger model with zero boundary conditions,  the interaction parameter $g=0.3$.}
\label{fig4}
\end{figure}

Electron density, calculated according to~(\ref{green}), equals
\begin{equation}
\rho(x)=\frac{N}{L}\left[1-2^{\frac{g}{2}}\sinh^g(\beta/2)\frac{\cos(2k_Fx-2f(x))}{[\cosh \beta-\cos(2\pi x/L)]^{g/2}}\right],
\label{denaver}
\end{equation}
with $f(x)$  being
\begin{equation}
 f(x)=\frac12\arctan\frac{\sin(2\pi x/L)}{e^{\beta}-\cos(2\pi x/L)}\;.
\end{equation}
Its Fourier-transform is presented in Fig.~(\ref{fig4}). The qualitative error of this result is that $\rho(q)$ does not vanish at $q>2k_F$, but, on the contrary, grows rapidly as $q\to 2k_F+0$. In other words, a whole branch appears in the region $q>2k_F$, which is absent in the exact solution.

Thus bosonization breaks down for the momenta close to $k_F$ and multiples of it. This scale corresponds to the mean inter-particle distance. Hence, the bosonization with zero boundary conditions incorrectly treats the short-ranged electron correlations, responsible for the formation of the ordered, Wigner-like state in 1D quantum dots. In the next section we demonstrate that the bosonization results describe the deformed Wigner crystal.

\section{Model of Wigner ordering}
\label{interpret}
Consider a simple model that takes into account the Wigner ordering from the very beginning. In this model, the many-electron wavefunction $\Phi(y_1,..,y_N)$ is represented by the Slater determinant
\begin{equation}
\Phi(y_1,..,y_N)\!=\!\dfrac{1}{\sqrt{N!}}
\begin{vmatrix}
 \phi(y_1\!-\!x_1)&\cdots&\phi(y_1\!-\!x_N)\\
 \vdots\\
 \phi(y_N\!-\!x_1)&\cdots&\phi(y_N\!-\!x_N)
\end{vmatrix},
\label{model_psi}
\end{equation}
built on the single-particle wavefunctions $\phi(x)=\pi^{-1/4}a^{-1/2}\exp(-x^2/2a^2)$, localized at positions $x_k$, $k=1\dots N$.  The wavefunction `width' $a$ is assumed to be smaller than the distance between the particles $L/N$, so the wavefunctions do not overlap and form an orthonormal set. 

The Fourier-transform of the density is
\begin{equation}
\rho(q)=e^{-\frac{\pi^2q^2a^2}{4L^2}}\sum_{k=1}^N \cos\frac{\pi qx_k}{L}\;.
\label{fourdenmod}
\end{equation}

The momentum distribution function $n(q)$ can be expressed through the electron Green function, similarly to Eq.~(\ref{mdf}),
\begin{equation}
n(q)=\int_0^L dxdy\,G(x,y)\psi_q^*(x)\psi_q(y)\;.
\label{pre_distr}
\end{equation}
The Green function $G(x,y)=\left\langle \psi^+(y)\psi(x)\right\rangle$ is related to the one-particle density matrix
\begin{equation}
\rho(x,y)=\int\Phi^*(y,z_2,..,z_N)\Phi(x,z_2,..,z_N)dz_2..dz_N\,,
\end{equation}
via $G(x,y)=N\rho(x,y)$.\cite{Mahan} Substitute this into Eq.~(\ref{pre_distr}) to get
\begin{equation}
n(q)=N\int d\mathbf{z}\left|\int dy\,\Phi^*(y,\mathbf{z})\psi_q(y)\right|^2\;,
\end{equation}
where $\mathbf{z}=(z_2,..,z_N)$. Using Eq.~(\ref{model_psi}), we finally obtain
\begin{equation}
\begin{split}
&n(q)=\sum_k\left|\int dy\,\phi^*(y-x_k)\psi_q(y)\right|^2 \\
&=4\sqrt{\pi}\frac{a}{L}e^{-\frac{\pi^2q^2a^2}{L^2}}\sum_{k=1}^N\sin^2\frac{\pi qx_k}{L}\;.
\label{mommod}
\end{split}
\end{equation}

First consider the case of a Wigner crystal, \textit{i.e.} when $x_k=(k-\frac12)\frac{L}{N}$. The Fourier-transform of the density $\rho(q)$ is non-zero only for multiples of $2k_F$,
\begin{equation}
\rho(n\cdot 2k_F)=(-1)^n N e^{-a^2 k_F^2 n^2}\;.
\end{equation}
The momentum distribution function $n(q)$ equals
\begin{equation}
n(q)=\frac{2\sqrt{\pi}aN}{L}\left[1-\sum_{m=1}^{\infty}(-1)^m\delta_{q,mN}\right]\exp\left(\!-\frac{a^2\pi^2}{L^2}q^2\right).
\end{equation}

\begin{figure}
\includegraphics[width=0.9\linewidth]{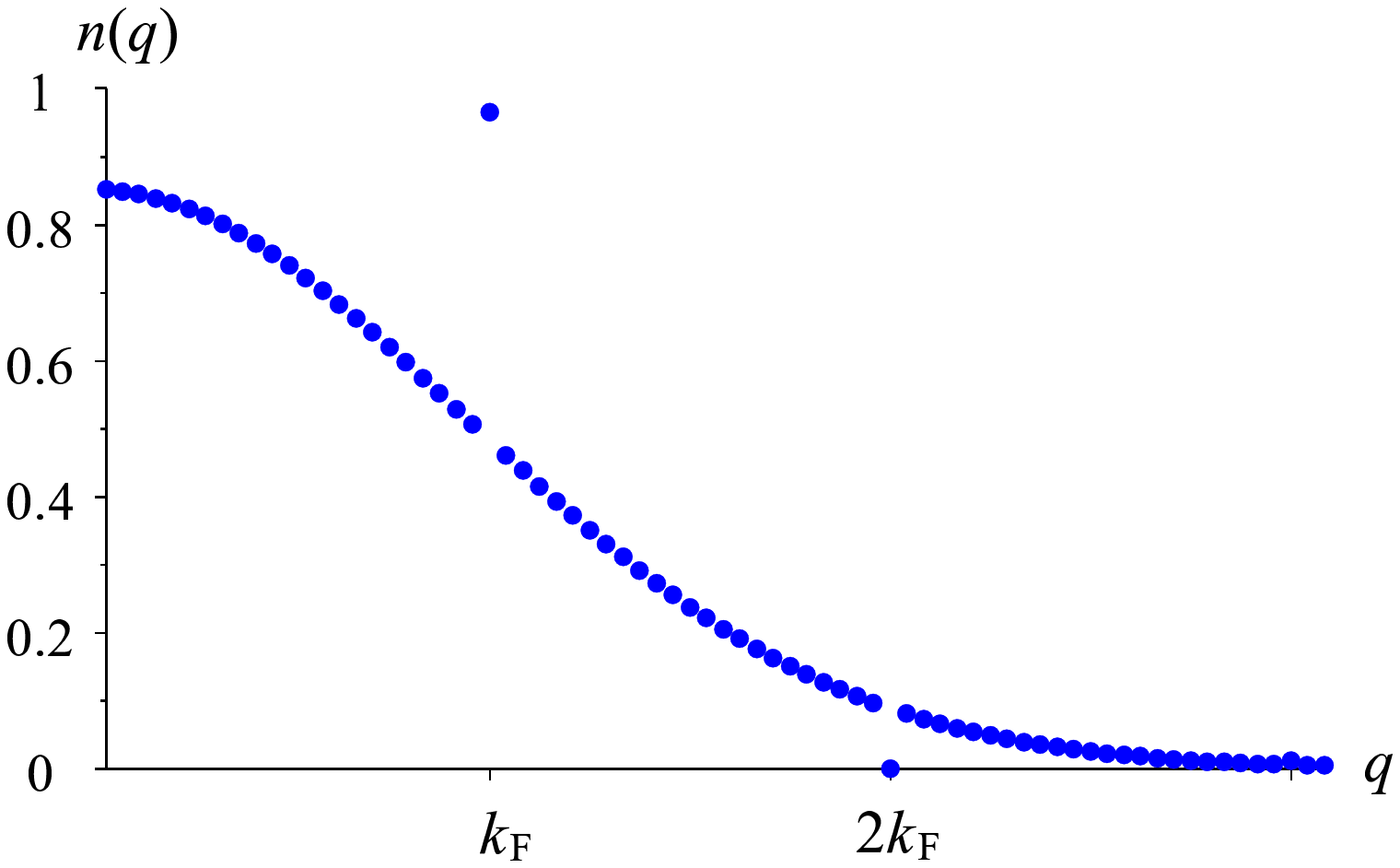}
\caption{Momentum distribution function in the model of localized electrons.}
\label{fig5}
\end{figure}

\begin{figure}
\includegraphics[width=0.9\linewidth]{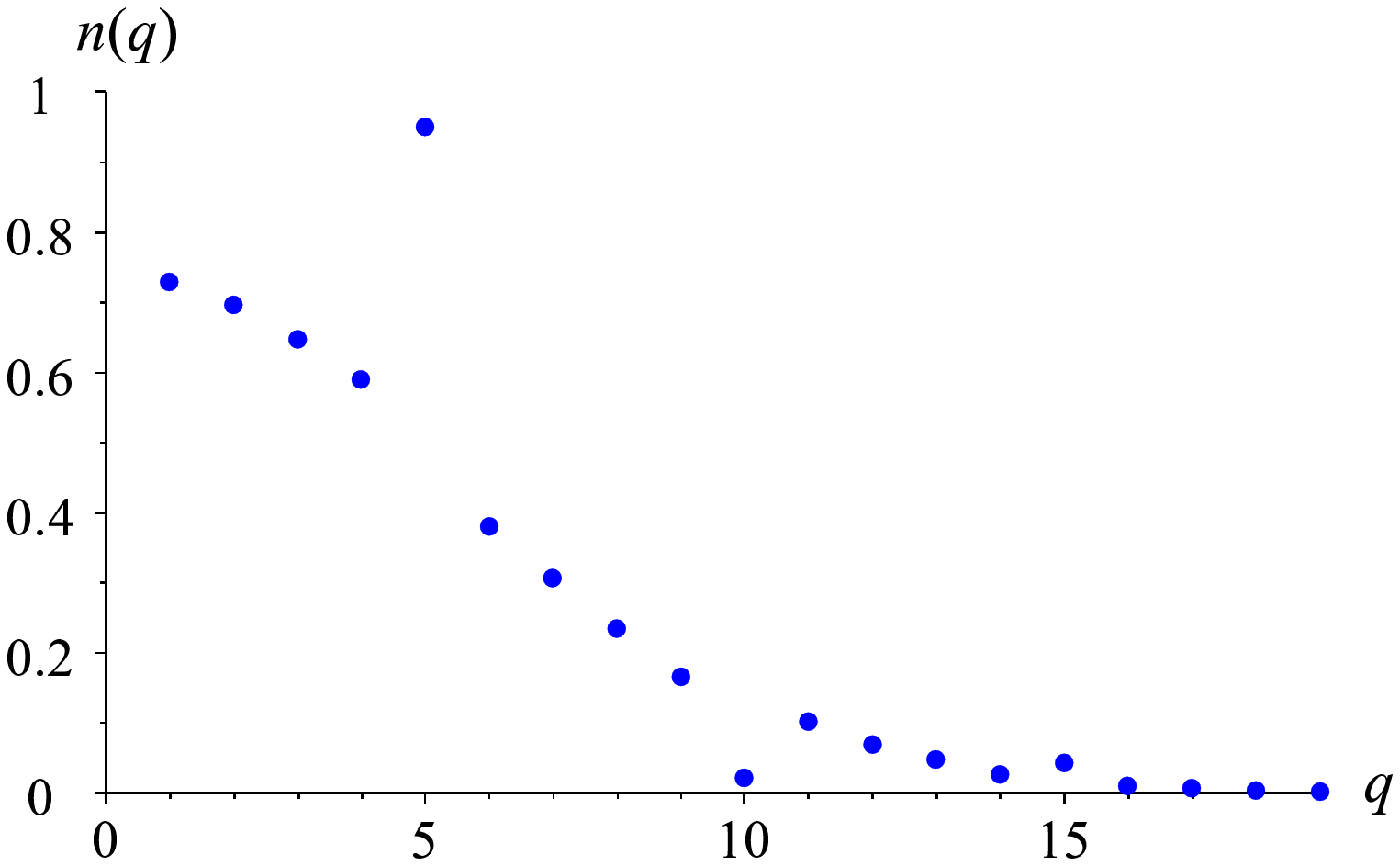}
\caption{Momentum distribution function, calculated by exact diagonalization for the system of $N=5$ electrons. The value of $q=5$ corresponds to $k_F$.}
\label{fig6}
\end{figure}

Fig.~\ref{fig5} shows that $n(q)$ is fully consistent with the results of exact diagonalization, including the $\delta$-singularity at the Fermi surface. Moreover, the additional $\delta$-singularities at multiples of $k_F$, predicted by the model, do exist in the exact solution for stronger interaction ($g\approx0.1$), see Fig.~\ref{fig6}.

This model proves that the origin of the singularities of $n(q)$ is the ordering of electrons in bounded 1D systems. In infinite 1D systems, the ordering is destroyed by fluctuations, and short-range electron correlations manifest themselves only in dynamic response to external perturbation.~\cite{Sablikov} If the system is finite, the boundaries pin the charge density waves, giving rise to Friedel oscillations, the amplitude of which is enhanced by \textit{e-e} interaction. This results in the increase of the weight of the state with $q=k_F$, which is reflected in the momentum distribution function, as well as in the high value of $2k_F$-harmonic of density. The conclusion about the important role of short-range electron correlations in bounded 1D systems is confirmed by the calculation of momentum distribution of electrons in a ring, \textit{i.e.} in a system without boundaries. In this case, the singularity of the momentum distribution disappears, as can be seen from Fig.~(\ref{fig3}).
\begin{figure}
\includegraphics[width=0.9\linewidth]{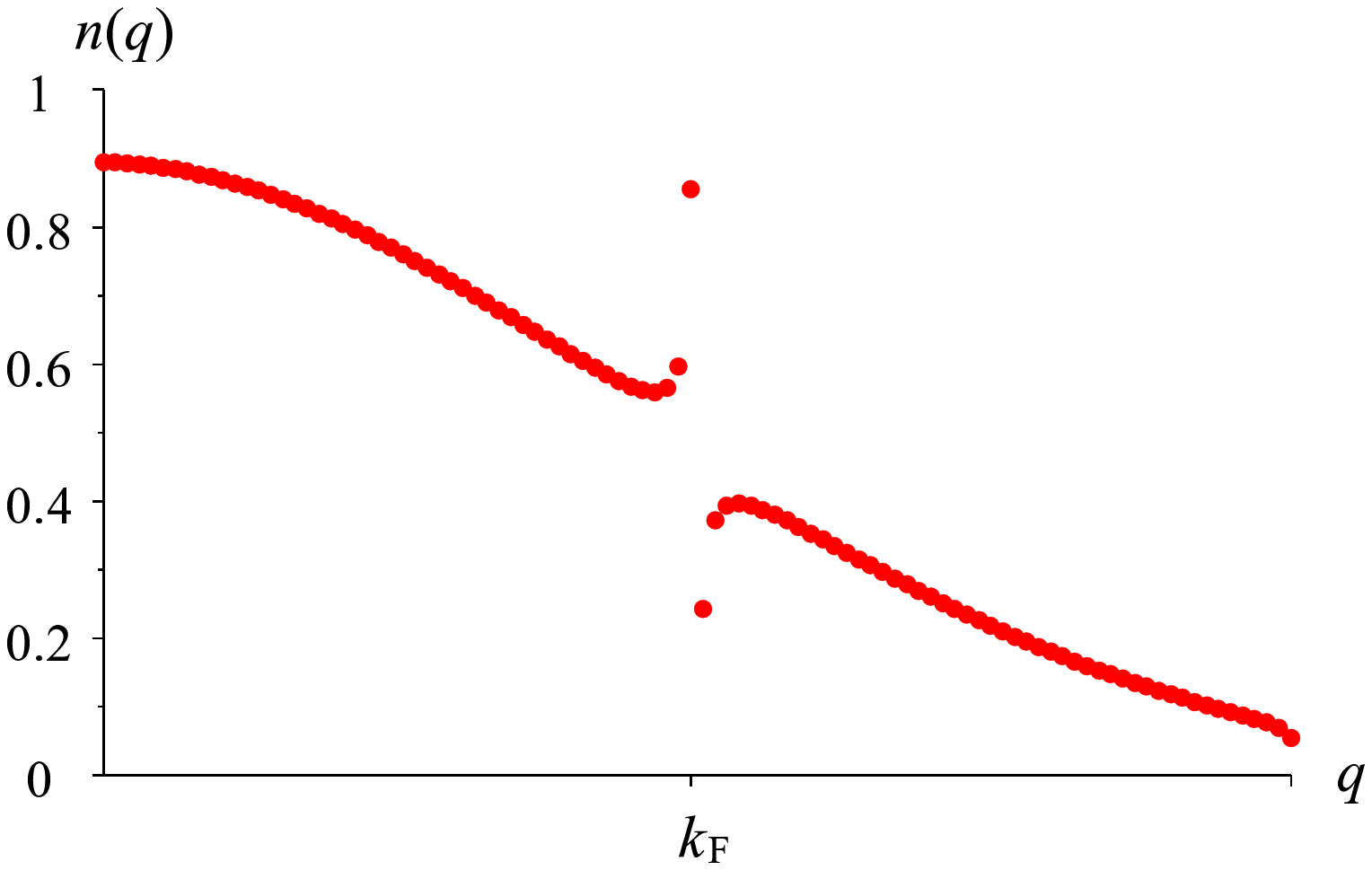}
\caption{Momentum distribution of the deformed Wigner crystal.}
\label{fig7}
\end{figure}
\begin{figure}
\includegraphics[width=0.9\linewidth]{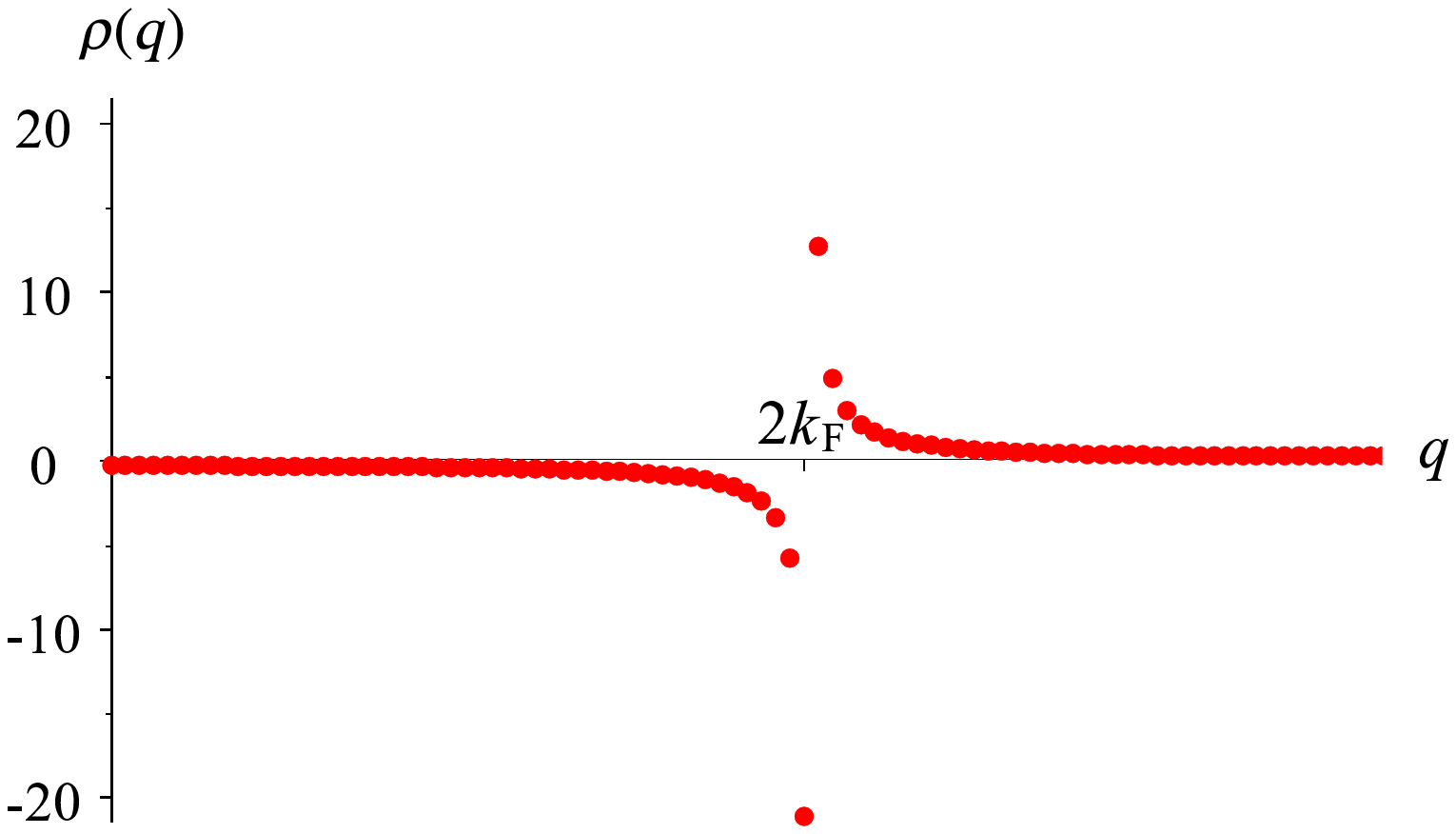}
\caption{The Fourier-transform of electron density for the deformed Wigner crystal.}
\label{fig8}
\end{figure}

Now let us squeeze electrons down to the center of the system by introducing a displacement $\delta x_k\propto -(k-\frac{N}{2})\frac{L}{N}$. The momentum distribution function and the density Fourier-spectrum, calculated according to (\ref{fourdenmod}),~(\ref{mommod}) for this case, are presented in Figs.~\ref{fig7},~\ref{fig8}. The curves are clearly identical to those from the previous section. This confirms our suggestion that bosonization describes the electron state that is deformed in comparison with the exact solution. The origin of such deformation is discussed in the next section.

\section{Bosonization with particle number conservation}
\label{bpnc}
\subsection{Formalism}
Let us briefly review the theory of the bounded Luttinger liquids, following~\cite{Fabrizio,Mattsson,Grioni}. Decompose the electron field operator
\begin{equation}
\psi(x)=\sum c_q\psi_q(x)
\label{field}
\end{equation}
into the fields $\psi_r(x)$ of the so-called $r$-fermions ($r=\pm$),
\begin{equation}
\psi(x)=e^{ik_Fx}\psi_+(x)+e^{-ik_Fx}\psi_-(x)\;,
\label{rel}
\end{equation}
where
\begin{equation}
\label{psir}
\psi_r(x)=-\frac{ir}{\sqrt{2L}}\sum_k c_ke^{irkx}\;.
\end{equation}
The fields are not independent, since
\begin{equation}
\psi_+(x)=-\psi_-(-x)\;,
\end{equation}
so we will deal with $\psi_+(x)$ alone. The latter has the property that
\begin{equation}
\psi_+(L)=\psi_+(-L)\;,
\end{equation}
\textit{i.e.} it satisfies periodic boundary conditions on the interval $[-L,L]$. Hence, it can be bosonized in a conventional way.

The main assumptions of bosonization are the linearization of $r$-fermion spectrum near the Fermi surface and its extrapolation to infinity. They allow one to solve the model exactly.

Consider the $r$-fermion density operator,
\begin{equation}
\rho(q)=\sum_k :c^+_{k+q} c_{k}:=\sum_k\left( c^+_{k+q}c_{k}-\delta_{q,0}\left\langle c^+_{k} c_{k}\right\rangle\right)
\end{equation}
for $q\ne0$. Zero harmonic $\rho(q=0)$ is the number of particles operator $\Delta N$. The density operator obeys the commutation relation
\begin{equation}
\left[\rho(q),\rho(-q')\right]=-\delta_{q,q'}\frac{qL}{\pi}\;,
\end{equation}
which allows one to introduce bosons
\begin{equation}
b^+_q=i\sqrt{\frac{\pi}{Lq}}\,\rho_q,\;q>0\;.
\end{equation}
The $r$-fermion field operator has the following boson representation,
\begin{equation}
\psi_+(x)=-\frac{i}{\sqrt{2\pi\alpha}}\,U\,e^{i\left(\phi(x)+\frac{\pi x}{L}\Delta N\right)}\;,
\label{bosident}
\end{equation}
where the bosonic phase $\phi$ equals
\begin{equation}
\phi(x)=\frac{i\pi}{L}\sum_{q\ne 0}\frac{e^{-iqx-\alpha|q|}}{q}\rho_q\;,
\label{bosphas}
\end{equation}
$U$ stands for the ladder operator, $\alpha$ denotes the ultraviolet cutoff, which by the order of magnitude equals $k_F^{-1}$.

The $r$-fermion spatial density is related to bosonic phase via
\begin{equation}
\rho_+(x)=:\psi^+_+(x)\psi_+(x):=\frac{\partial_x\phi}{2\pi}+\frac{\Delta N}{2\pi}\;,
\label{bos_repres}
\end{equation}
and $\rho_-(x)=\rho_+(-x)$.

To obtain the density operator of real electrons, one has to express $\psi(x)$ in terms of $\psi_r(x)$. Using Eq.~(\ref{rel}), one gets
\begin{equation}
\rho(x)=\rho_\mathrm{lw}(x)+\rho_\mathrm{CDW}(x)\;,
\end{equation}
where the first term is the long-wave component of the density,
\begin{equation}
\rho_\mathrm{lw}(x)=\rho_+(x)+\rho_-(x)\;,
\end{equation}
and the second term is the charge density wave (CDW) component
\begin{equation}
\rho_{\mathrm{CDW}}(x)=e^{-2ik_Fx}\psi_+^+(x)\psi_-(x)+h.c.
\label{rhocdw}
\end{equation}
which describes short-range electron correlations. In terms of bosonic phase, the electron density operator equals
\begin{equation}
\rho(x)=\frac{k_F}{\pi}-\frac{1}{\pi}\frac{\partial \varphi}{\partial x}-\frac{k_F}{\pi}\cos(2k_Fx-2\varphi(x)-2f(x)),
\label{real_density}
\end{equation}
where $\varphi(x)=\frac12(\phi(-x)-\phi(x))$, and function $f(x)$ is an additional phase due to zero boundary conditions,
\begin{equation}
 f(x)=\frac{1}{4i}[\phi(x),\phi(-x)]=\frac12\arctan\frac{\sin(2\pi x/L)}{e^{\beta}-\cos(2\pi x/L)},
\end{equation}
with dimensionless cutoff $\beta=\pi\alpha/L\approx N^{-1}$.

The kinetic energy
\begin{equation}
H_0=v_F\sum_k k:c_k^+c_k:
\end{equation}
is bosonized using Kronig's identity to give
\begin{equation}
H_0=v_F\sum_{q>0} q:b_q^+b_q:+\frac{\pi v_F}{2L}(\Delta N)^2\;.
\end{equation}

The interaction part of the Hamiltonian equals
\begin{equation}
V=\frac12\int_0^L dxdy\, \rho(x)V(x-y)\rho(y)\;,
\label{interactpart}
\end{equation}
where $\rho(x)$ is given by Eq.~(\ref{real_density}). In the model considered,~\cite{Fabrizio,Mattsson,Grioni} the interaction operator is simplified by retaining only the direct
\begin{equation}
V_d=\frac12\int_0^L dxdy\, \left( \rho_+(x)\rho_+(y)+\rho_-(x)\rho_-(y)\right) V(x-y)\;
\end{equation}
and
cross terms
\begin{equation}
V_c=\frac12\int_0^L dxdy\, \left( \rho_+(x)\rho_-(y)+\rho_-(x)\rho_+(y)\right) V(x-y)\;.
\end{equation}
The cross term contains the non-local contribution  $\rho_+(x)\rho_-(y)=\rho_+(x)\rho_+(-y)$, which makes the diagonalization of the Hamiltonian impossible, unless the interaction potential is assumed short-ranged, that is of the form $V\delta(x-y)$. Under this assumption, the direct and cross terms transform into
\begin{align}
V_d&=\frac{V}{2}\int_{-L}^{L} dx\,  \rho_+^2(x)\;,\\
V_c&=\frac{V}{2}\int_{-L}^{L} dx\,  \rho_+(x)\rho_+(-x)\;.
\end{align}

Combining the terms, and using bosonic representation~(\ref{bos_repres}) of the density operator, we arrive at the following model Hamiltonian:
\begin{equation}
\begin{split}
H&=H_0+V_d+V_c=(v_F+\frac{V}{2\pi})\sum_{q>0}q\left( b_q^+b_q+\frac12\right)-\\ 
\frac{V}{4\pi}&\sum_{q>0} q\left( b_q^+b_q^+ +b_q b_q\right)+
\frac{\pi}{2L}(v_F+\frac{V}{\pi})(\Delta N)^2.
\end{split}
\end{equation}
The Hamiltonian is diagonalized by Bogoliubov transformation
\begin{equation}
\tilde{b}_q=e^{iS}b_qe^{-iS}\;,
\end{equation}
where
\begin{equation}
S=\frac{i}{2}\sum_{q>0}\xi_q\left( b_q^+b_q^+-b_q b_q \right)\;.
\end{equation}
The Hamiltonian is diagonal if $e^{2\xi} =(1+V/\pi v_F)^{-1/2}$. The interaction parameter $g\equiv e^{2\xi}$ belongs to $(0,1)$ for repulsive interaction, and equals unity for free electrons. The diagonal form of the Hamiltonian is
\begin{equation}
H=\sum_{q>0}qv(q)\,\tilde{b}_q^+\tilde{b}_q+\frac{\pi}{2L}v_N(\Delta N)^2\;,
\label{quadr_ham}
\end{equation}
with
\begin{equation}
v(q)=\frac{v_F+V/2\pi}{\cosh 2\xi},\;v_N=v_F+\frac{V}{\pi}\;.
\end{equation}
The bosonic representation~(\ref{bosident}) of the field operator, the density operator~(\ref{real_density}) and quadratic Hamiltonian~(\ref{quadr_ham}) are sufficient to obtain the Green function~(\ref{green}) with all the ensuing problems for the momentum distribution~(\ref{mdf}) and density~(\ref{denaver}). The root of these problems lies in the density operator~(\ref{real_density}), commonly used in the bosonization approach. 

\subsection{Density operator}
The bosonization approach as presented above violates the particle number conservation. To see this, just notice that the integral of the density fluctuation (the second term in Eq.~(\ref{denaver})) over the system length is not zero. This is a well-known problem, which exists not only in the case of zero boundary conditions, but also in a standard bosonization on the ring.~\cite{Sablikov} The problem arises at the level of the electron density operator~(\ref{real_density}), the CDW component of which does not conserve the number of particles in an isolated system. The physical reason of the violation of the particle number conservation is that the CDW component of~(\ref{real_density}) includes the response of the infinite positron sea, which is not completely eliminated when using the approximate relation~(\ref{rel}).

In the case of zero boundary conditions, the situation is reacher because now there appear problems even with the long-wave component of the density. Indeed, Eq.~(\ref{denaver}) of the common Luttinger liquid theory does not give a correct transition to the case of non-interacting electrons in the box. The density of free electrons
\begin{equation}
\rho_\mathrm{free}(x)=\frac{N}{L}+\frac{1}{2L}-\frac{\sin(2k_F+\frac{\pi}{L})x}{2\sin \frac{\pi x}{L}}
\label{denfree}
\end{equation}
contains an additional term $1/2L$, missing in~(\ref{denaver}). This term, being integrated over the length of the system, gives an extra charge of $e/2$.

Thus the microscopic theory leads to the density operator that violates the particle number conservation. We will obtain the correct operator, following the harmonic-fluid approach by Haldane.~\cite{Haldane_thin}

Introduce the phase $\theta(x)$ that increases by $\pi$ each time $x$ passes the location $x_k$ of a $k$-th particle. The particle density operator~(\ref{pardenop}) becomes
\begin{equation}
\rho(x) = \partial_x\theta\sum_{k=1}^N\delta(\theta(x)-k\pi) = \frac{\partial_x \theta}{\pi} \sum_{m=-\infty}^{\infty}e^{2im\theta(x)}\;.
\label{hal}
\end{equation}
According to Eq.~(\ref{real_density}),
\begin{equation}
\theta(x) = k_Fx - \varphi(x) - f(x)\;. 
\end{equation}
Thus the density operator takes the form
\begin{equation}
\begin{split}
&\rho(x)=\frac{k_F}{\pi}-\frac{\partial_x\varphi}{\pi}-\frac{\partial_x f}{\pi}\\
&-(\frac{k_F}{\pi}-\frac{\partial_x\varphi}{\pi}-\frac{\partial_x f}{\pi})\cos(2k_Fx-2\varphi(x)-2f(x)),
\label{denfin}
\end{split}
\end{equation}
where we retain only $m=0,\pm1$ harmonics, and halve the amplitude of the CDW component to obtain the correct transition to the non-interacting case.~\cite{Sablikov}

The density operator of Eq.~(\ref{denfin}) has the form of a full differential, which guarantees that the integral of the density fluctuation over the length of the system is zero, \textit{i.e.} the number of particles is conserved. The long-wave part of the operator contains an additional term $-\partial_x f/\pi$ that gives the $1/2L$ component, missing in~(\ref{denaver}), since in the bulk of the system $f(x)\approx\frac{\pi}{4}-\frac{\pi x}{2L}$.

\subsection{Hamiltonian and observables}
The interaction part~(\ref{interactpart}) of the Hamiltonian, having been calculated with the density operator of Eq.~(\ref{denfin}), gets additional terms of the form~\cite{footnote2}
\begin{equation}
H_1=\frac{V}{2\pi^2}\int_0^Ldx\,\partial_x\varphi\partial_xf=iV\sum_{q>0}\sqrt{\frac{gq}{8\pi L}}(\tilde{b}^+_{2q}-\tilde{b}_{2q}).
\end{equation}
By shifting bosons 
\begin{equation}
d_{2q}=\tilde{b}_{2q}+i\frac{V}{4v_F}\sqrt{\frac{g^3}{2\pi qL}}
\end{equation}
we obtain a diagonalized full Hamiltonian $H+H_1$,
\begin{equation}
H+H_1=\sum_{q>0}qv(q) d_q^+d_q\;.
\end{equation}
The bosonic phase~(\ref{bosphas}) is transformed into $\phi(x)=\phi_0(x)+\phi_1(x)$, where $\phi_0(x)$ is linear in new bosons,
\begin{equation}
\phi_0(x)\!=\!\sum_{q>0}\sqrt{\frac{\pi}{qL}}\left[(ce^{iqx}\!-\!se^{-iqx})d_q+(ce^{-iqx}\!-\!se^{iqx})d_q^+\right]\;,
\end{equation}
with $c$ and $s$ being, respectively, $\cosh\xi$ and $\sinh\xi $. The function $\phi_1(x)$ is the new phase, specific to the case of zero boundary conditions, $\phi_1(x)=Af(x)$, $A=Vg^2/2\pi v_F$.

As a result, the field operator~(\ref{bosident}) acquires the factor of $\exp(iAf(x))$, and the Green function~(\ref{green}) transforms as 
\begin{equation}
G_\mathrm{new}(x,y)=G_+(x,y)e^{iA(f(x)-f(y))}\;.
\label{grnew}
\end{equation}

The average value of the density equals
\begin{equation}
\begin{split}
\langle\rho(x)\rangle&=\frac{k_F}{\pi}-(1-A)\frac{\partial_x f}{\pi}\\
&-\frac{\sinh^g(\beta/2)}{2^{1-g/2}\pi}\frac{\partial}{\partial x}\frac{\sin(2k_Fx-2(1-A)f(x))}{[\cosh\beta-\cos\frac{2\pi x}{L}]^{g/2}}\;.
\label{correctdensity}
\end{split}
\end{equation}

Note that the additional phase $(1-A)f(x)$, which appears in Eq.~(\ref{correctdensity}), changes the period of density oscillations. Since $\partial_x f<0$ everywhere in the system except narrow regions near the ends, the Wigner crystal is squeezed by the boundaries. The deformation is determined by the coefficient $(1-A)$. In the common Luttinger liquid theory $A=0$. Our approach with corrected expression (\ref{denfin}) for the density operator gives $A>0$. Hence, restoring total neutrality results in the reduction of the Wigner crystal compression.

The distribution function $n(q)$ and the Fourier spectrum of the density calculated with the use of the above expression for $A$ coincide with the numerical calculations quite well. However this formula is justified for weak interaction ($1-g\ll 1$). We propose a generalized expression for any $g$ taking into account that $A$ is known for three limiting cases: 

i) For $g=1$, $A=0$ to provide the correct transition to the case of non-interacting electrons.

ii) For weak interaction, $A$ should be proportional to $V/2\pi v_F$, in agreement with our model.

iii) For strong interaction $g\to 0$, $A\to 1$ to provide the transition to the limiting case of the Wigner crystal with strictly periodic density.

\begin{figure}
\includegraphics[width=0.9\linewidth]{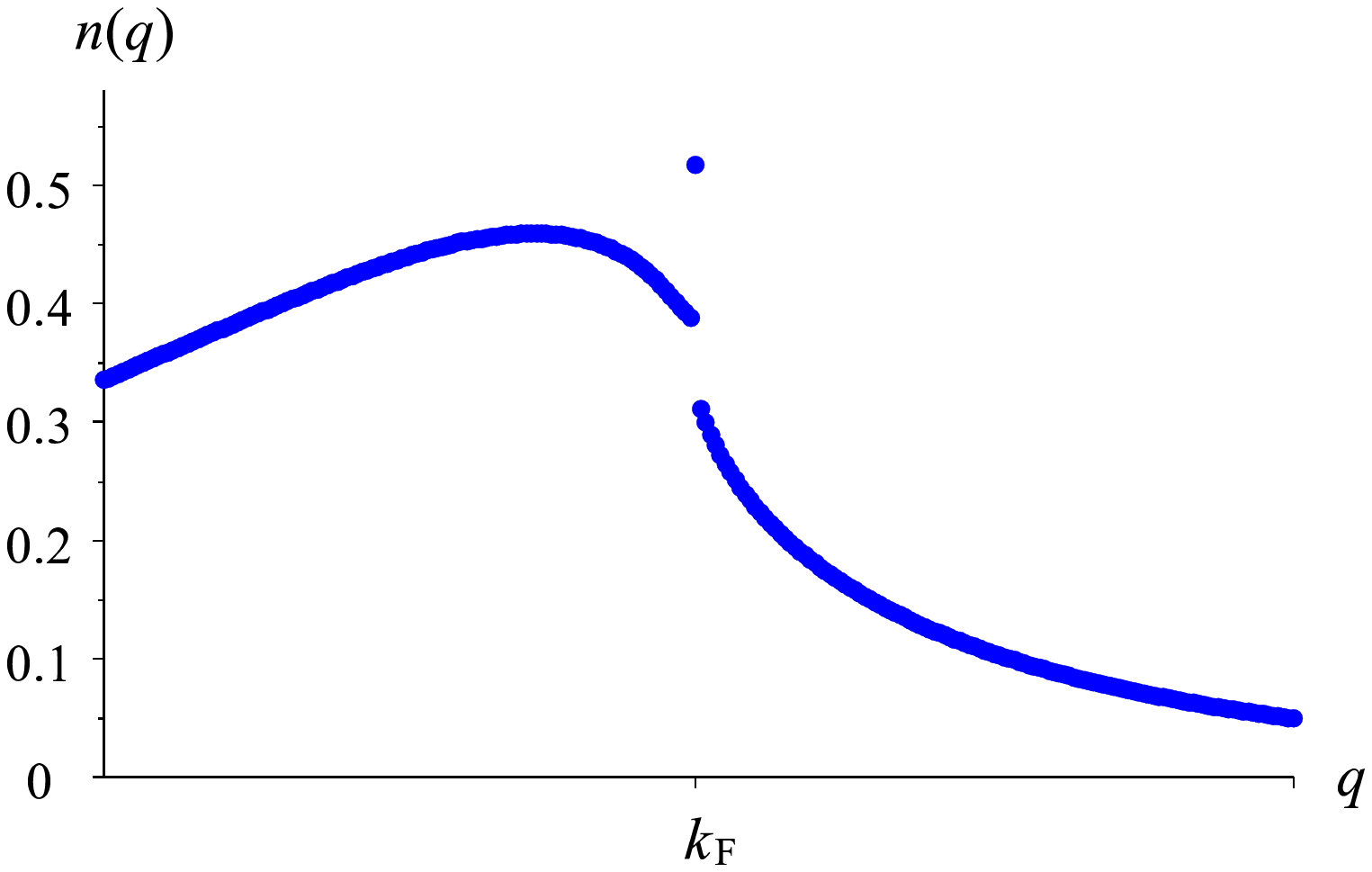}
\caption{Momentum distribution function, the violation of the particle number conservation is fixed.}
\label{fig9}
\end{figure}

\begin{figure}
\includegraphics[width=0.9\linewidth]{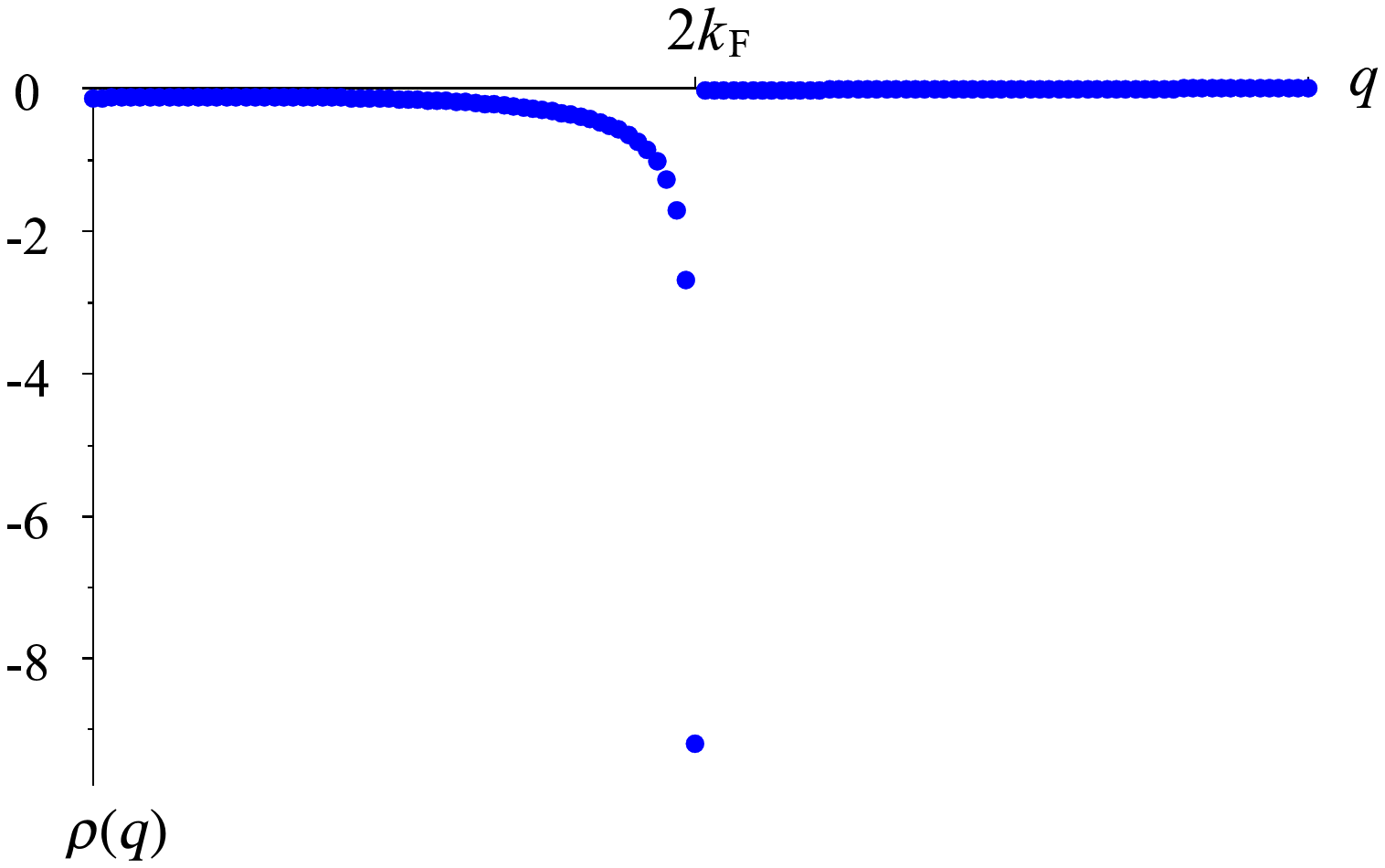}
\caption{The Fourier-transform of electron density, the violation of the particle number conservation is fixed.}
\label{fig10}
\end{figure}

The simplest choice of $A=1-g$ satisfies these requirements and proves to be highly successful, as is demonstrated below. Figs.~\ref{fig9},~\ref{fig10} show the momentum distribution function and the density Fourier-transform, calculated according to Eqs.~(\ref{grnew}),~(\ref{correctdensity}). They are seen to agree nicely with the exact results of section~\ref{eds}.


\begin{figure}
\includegraphics[width=1.0\linewidth]{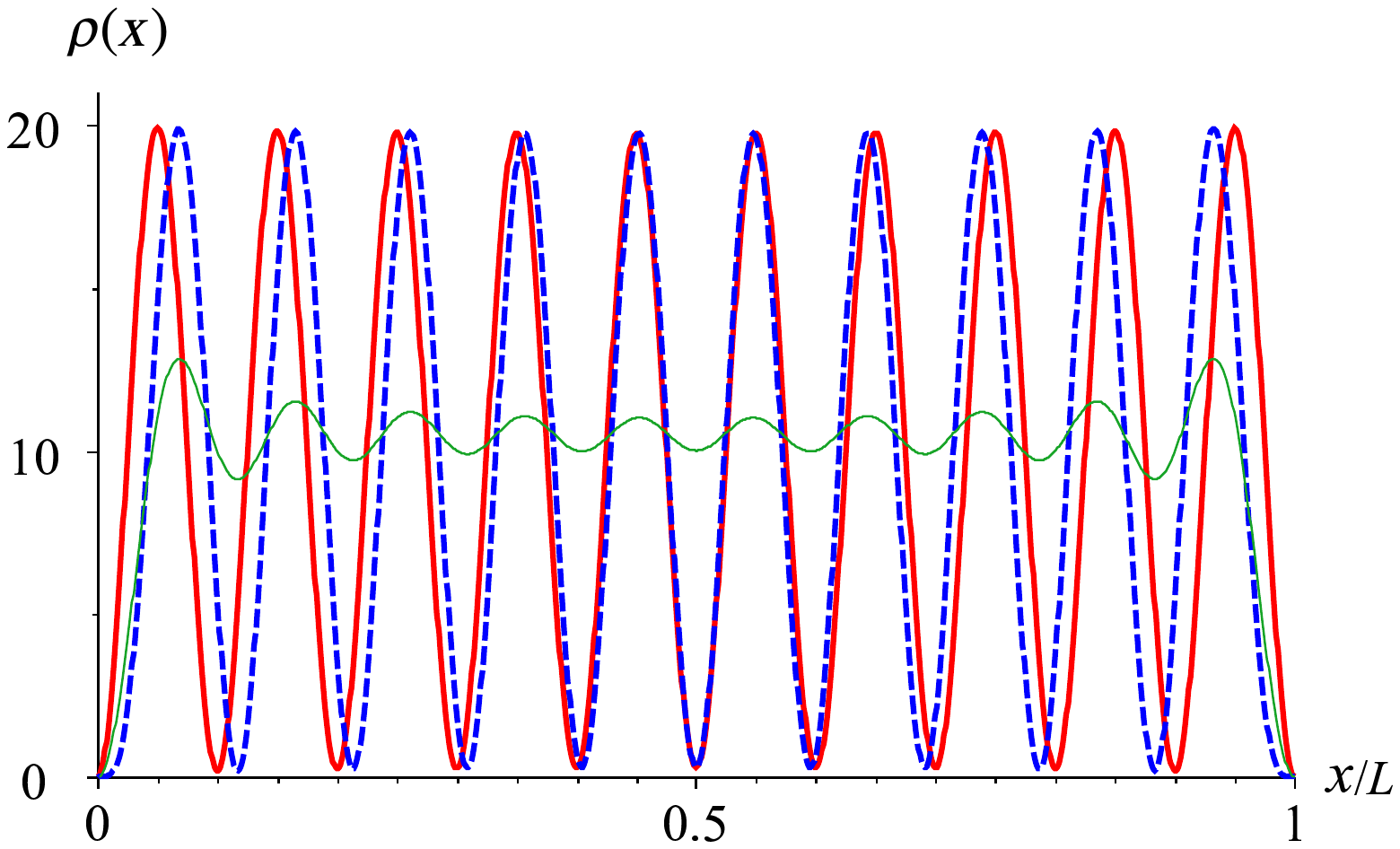}
\caption{The spatial distribution of the density, calculated according to Eqs.~(\ref{denaver}) (dashed line), Eq.~(\ref{denfree}) (thin solid line), Eq.~(\ref{correctdensity}) (solid line).}
\label{fig11}
\end{figure}

Fig.~\ref{fig11} shows the spatial distribution of the density, calculated according to Eqs.~(\ref{denaver}),~(\ref{denfree}),~(\ref{correctdensity}). It is seen that according to the standard bosonization the electron density maxima almost coincide with those of free electrons, even for strong interaction. In contrast, according to Eq.~(\ref{correctdensity}) the electron locations are shifted towards the periodic positions as Wigner ordering prescribes at strong interaction. Thus the bosonization approach that does not respect the particle number conservation leads to the picture of the state with electrons squeezed to the center of the system.

\section{Concluding remarks}
In this work, the ground state of interacting electrons in a 1D quantum dot was investigated using the exact diagonalization. An unexpected $\delta$-like singularity was discovered in the momentum distribution function at the Fermi energy. A threshold behavior was found for the spatial Fourier-spectrum of electron density, with the step at $2k_F$. These effects are stable against the change in the system length, interaction radius, the number of electrons, and the interaction strength. Thus we suggest that they are inherent in finite 1D electronic systems. We proposed a simple model which proved that these effects originate from the formation of the Wigner molecule in a 1D quantum dot. Comparison of exact results with the Luttinger model with zero boundary conditions shows that the latter does not correctly describe both the momentum distribution near the Fermi energy and the density Fourier-spectrum. The problem is that bosonization overestimates the deformation of the Wigner crystal caused by the boundaries by introducing the excessive positive charge into the 1D system, which attracts electrons to the center of the system. This is a result of using the density operator that violates the number of particles conservation. We derived the density operator devoid of the mentioned shortcoming, corrected the Hamiltonian, and calculated the observables to find a nice agreement with the exact results.

We emphasize that our results are specific to finite 1D systems with boundaries. The above features of the momentum distribution function and density Fourier-spectrum are both kept with increasing the system length. As far as the spectrum remains discrete, the $\delta$-singularity of $n(q)$ refers to a single point $q=k_F$. Formally, the features survive even at $L\to \infty$. However, this is not equivalent to the thermodynamic limit, which also requires that dephasing processes be taken into account. In the correct thermodynamic limit the features disappear together with the boundary effects.

\acknowledgments
This work was supported by Russian Foundation for Basic Research (project No. 05-02-16854), Russian Academy of Sciences (programs “Quantum Nanostructures” and “Strongly Correlated Electrons in Semiconductors, Metals, Superconductors, and Magnetic Materials”), RF Ministry of Education and Science, and Russian Science Support Foundation.

\appendix
\section{Hamiltonian matrix elements}
\label{meth}
The Hamiltonian matrix elements
\begin{equation}
H_{\mathbf{p}_1\mathbf{p}_2}=\int \Phi^*_{\mathbf{p}_1}H\Phi_{\mathbf{p}_2}\,dx_1..dx_N
\label{matrix_elements}
\end{equation}
are calculated by substituting Eqs.~(\ref{Hamiltonian}),~(\ref{many_particle_wf}) into Eq.~(\ref{matrix_elements}), expanding the Slater determinants, and using the orthogonality property of single-particle functions~(\ref{one_particle_wf}) during integration.~\cite{Slater1,Condon,Slater2,Lowdin}

The matrix element of the kinetic energy equals
\begin{equation}
T_{\mathbf{p}_1\mathbf{p}_2}=\frac{\hbar^2 \pi^2}{2mL^2}\sum_{i=1}^{N}k_i^2\,\delta_{\mathbf{p}_1\mathbf{p}_2}\;.
\end{equation}

For the matrix elements of the pair \textit{e-e} interaction, four situations are possible, depending on $\mathbf{p}_1$ and $\mathbf{p}_2$.
\begin{enumerate}
\item If $\mathbf{p}_1=\mathbf{p}_2=(\alpha_1,..,\alpha_N)$, \textit{i.e.} diagonal matrix elements are calculated, then
\begin{equation}
 V_{\mathbf{p}_1\mathbf{p}_1}=\sum_{i>j=1}^{N}V_{\alpha_i\alpha_j,\alpha_i\alpha_j}\;,
\end{equation}
where
\begin{equation}
\begin{split}
&\quad V_{\alpha_i\alpha_j,\alpha_k\alpha_l}=\\
&\quad \int \phi^*_{\alpha_i\alpha_j}(x_1,x_2)V(x_1-x_2)\phi_{\alpha_k\alpha_l}(x_1,x_2)dx_1dx_2,
\end{split}
\label{interaction_matrix}
\end{equation}
and
\begin{equation}
\phi_{\alpha_i\alpha_j}(x_1,x_2)=\dfrac{1}{\sqrt{2!}}
\begin{vmatrix}
\psi_{\alpha_i}(x_1)&\psi_{\alpha_j}(x_1)\\
\psi_{\alpha_i}(x_2)&\psi_{\alpha_j}(x_2) 
\end{vmatrix}\;.
\end{equation}
\item If $\mathbf{p}_1$ and $\mathbf{p}_2$ are identical except two states, numbered $k_1$ and $k_2$, respectively, \textit{i.e.} if $\mathbf{p}_1=(\alpha_1,..\alpha_{k_1},..\alpha_N)$, $\mathbf{p}_2=(\alpha_1,..\beta_{k_2},..\alpha_N)$, then 
\begin{equation}
V_{\mathbf{p}_1\mathbf{p}_2}=(-1)^{k_1+k_2}\sum_{\substack{i=1 \\ i\ne k_1}}^{N}V_{\alpha_{k_1}\alpha_i,\beta_{k_2}\alpha_i}\;,
\end{equation}
\item If $\mathbf{p}_1$ and $\mathbf{p}_2$ are identical except four states, numbered $k_i$, $i=1\dots4$, \textit{i.e.} if $\mathbf{p}_1=(\alpha_1,..\alpha_{k_1},..\alpha_{k_2},..\alpha_N)$, $\mathbf{p}_2=(\alpha_1,..\beta_{k_3},..\beta_{k_4},..\alpha_N)$, then
\begin{equation}
V_{\mathbf{p}_1\mathbf{p}_2}=(-1)^{\sum \limits_{i=1}^4 k_i} V_{\alpha_{k_1}\alpha_{k_2},\beta_{k_3}\beta_{k_4}}\;.
\end{equation}
\item If $\mathbf{p}_1$ and $\mathbf{p}_2$ contain more than four not coincident states $\alpha_i$, then
\begin{equation}
V_{\mathbf{p}_1\mathbf{p}_2}=0\;.
\end{equation}
\end{enumerate}

Matrix elements of the interaction between electrons and positive background are as follows.
\begin{enumerate}
 \item If $\mathbf{p}_1=\mathbf{p}_2=(\alpha_1,..,\alpha_N)$, then
\begin{equation}
 U_{\mathbf{p}_1\mathbf{p}_1}=\sum_{i=1}^{N}f_{\alpha_i\alpha_i}\;,
\end{equation}
where
\begin{equation}
f_{\alpha_i\alpha_k}=\int \psi^*_{\alpha_i}(x)U(x)\psi_{\alpha_k}(x)dx\;.
\label{matrix_ion}
\end{equation}
\item If $\mathbf{p}_1$ and $\mathbf{p}_2$ are identical except two states, numbered $k_1$ and $k_2$, respectively, \textit{i.e.} if $\mathbf{p}_1=(\alpha_1,..\alpha_{k_1},..\alpha_N)$, $\mathbf{p}_2=(\alpha_1,..\beta_{k_2},..\alpha_N)$, then
\begin{equation}
 U_{\mathbf{p}_1\mathbf{p}_2}=f_{\alpha_{k_1}\beta_{k_2}}\;.
\end{equation}
\item Otherwise, $U_{\mathbf{p}_1\mathbf{p}_2}=0$.
\end{enumerate}

The matrix element of the pair \textit{e-e} interaction~(\ref{interaction_matrix}) equals
\begin{equation}
\begin{split}
V_{lj,km}&=\dfrac{Ve^2d}{2\epsilon L^2}[f(l,j,k,m)-f(l,j,k,-m)-\\
&f(l, j, -k, m)+f(l, j, -k, -m)-f(l, -j, k, m)+\\
&f(l, -j, k, -m)+f(l, -j, -k, m)-f(l, -j, -k, -m)-\\
&f(j, l, k, m) + f(j, l, k, -m) + f(j, l, -k, m)-\\
&f(j, l, -k, -m)+f(j, -l, k, m)-f(j, -l, k, -m)-\\
&f(j, -l, -k, m) + f(j, -l, -k, -m)],
\end{split}
\end{equation}
where $f(l,n,k,m)=g(l+k,n+m)$, and $g(p,q)=0$ for odd $(p+q)$, while for even $(p+q)$
\begin{equation}
\begin{split}
g(p,q)=(e^{-a}(-1)^p-1)\frac{1+\frac{\pi^2}{a^2}pq}{(1+\frac{\pi^2}{a^2}p^2)(1+\frac{\pi^2}{a^2}q^2)}+&\\
\frac{a}{1+\frac{\pi^2}{a^2}p^2}\delta_{p,-q}&\;,
\end{split}
\end{equation}
with $a=L/d$. Matrix element~(\ref{matrix_ion}) equals
\begin{equation}
f_{ln}=N\frac{Ve^2d}{\epsilon L^2}(g(l+n,0)-g(l-n,0))\;.
\end{equation}


\begin{thebibliography}{99}
\bibitem{footnote1}
Since bosonization is applicable only for low-energy excitations, the results for $n(q)$ can only be considered in the vicinity of the Fermi surface, at $|k_F-q|\ll k_F$. The suppression of $n(q)$ at $k_F-q\approx k_F$ is the consequence of ultraviolet cutoffs, performed in the theory to exclude the high-energy contribution of the positron sea. Cutoffs should not affect the low-energy region near $k_F$.
\end{thebibliography}
\end{document}